\documentclass[aps,pra,reprint,superscriptaddress,amsmath,amssymb]{revtex4-2}
\usepackage{graphicx}
\usepackage[colorlinks=true,bookmarks=false,citecolor=blue,urlcolor=blue]{hyperref} 
\usepackage[table,xcdraw,dvipsnames]{xcolor}
\usepackage[normalem]{ulem}

\begin{document}
\title{Searching for Axions on a Higher Note: \\ Third-Harmonic Generation from Colliding High-Intensity Laser Beams}
\author{Ashis Paul} 
\affiliation{Department of Physical and Chemical Sciences, University of L'Aquila, Via Vetoio, 67100 L'Aquila, Italy}

\author{Leone di Mauro Villari}
\affiliation{Department of Physical and Chemical Sciences, University of L'Aquila, Via Vetoio, 67100 L'Aquila, Italy}

\author{Davide Tedeschi}
\affiliation{Department of Physical and Chemical Sciences, University of L'Aquila, Via Vetoio, 67100 L'Aquila, Italy}

\author{Carino Ferrante}
\affiliation{CNR-SPIN, c/o Dip.to di Scienze Fisiche e Chimiche, Via Vetoio,  L'Aquila 67100, Italy}

\author{Luca Di Luzio} 
\email{luca.diluzio@pd.infn.it}
\affiliation{INFN Sezione di Padova, Via Francesco Marzolo 8, 35131 Padova, Italy}

\author{Andrea Marini} 
\email{andrea.marini@univaq.it}
\affiliation{Department of Physical and Chemical Sciences, University of L'Aquila, Via Vetoio, 67100 L'Aquila, Italy}
\affiliation{CNR-SPIN, c/o Dip.to di Scienze Fisiche e Chimiche, Via Vetoio,  L'Aquila 67100, Italy}

\begin{abstract}
We propose a new laboratory strategy to generate and detect axion-like particles via third-harmonic generation induced by two non-collinear, polarised high-intensity laser beams of peak intensity of the order of $10^{24}\,\mathrm{W/cm^2}$, where the third-harmonic signal is generated by the axion field.
Starting from the axion-modified Maxwell equations, we analytically derive the axion-induced third-harmonic field, and show that by using state-of-the-art petawatt laser facilities, a detectable signal can be obtained over a broad range of axion masses and couplings. A key feature of the setup is that the axion-photon conversion rate can be resonantly enhanced by tuning the angle between the two beams through a mechanism that does not depend on the physical volume of the apparatus. The proposed configuration may therefore probe an unexplored region of axion parameter space and pave the way for next-generation high-power laser-based axion searches.

\end{abstract}
\maketitle

\section{Introduction}

The axion was originally introduced as a dynamical solution to the strong (charge-parity) CP problem of quantum chromodynamics (QCD)
\cite{Peccei:1977hh,Peccei:1977ur,Weinberg:1977ma,Wilczek:1977pj}.
Remarkably, it was soon realised that axions can also constitute an excellent dark matter candidate~\cite{Preskill:1982cy,Abbott:1982af,Dine:1982ah}. More generally, axion-like particles arise naturally in many extensions of the Standard Model, including
theories with broken global symmetries \cite{Chikashige:1980ui}
and string compactifications \cite{Witten:1984dg}. Their low-energy phenomenology is often accounted for through the axion-photon interaction Lagrangian density 
\begin{equation}
{\cal L}_{\rm int}
=
-\frac{1}{4}g_{a\gamma} a F_{\mu\nu}\tilde F^{\mu\nu}
=
g_{a \gamma} a\, {\bf E}\cdot{\bf B} ,
\end{equation}
which describes the axion--photon interaction, with $g_{a\gamma}$
given in natural units and quoted in ${\rm GeV}^{-1}$.
Note that the interaction Lagrangian density depends on ${\bf E}\cdot{\bf B}$, where ${\bf E}$ and ${\bf B}$ are electric and magnetic fields, and thus vanishes for electromagnetic (EM) plane waves. However, one can obtain a non-zero interaction Lagrangian by non-collinear excitation schemes, which is the subject of this paper, see below, or more traditionally through EM propagation in static electric or magnetic fields. This interaction provides one of the most direct experimental handles on axions and has motivated a broad experimental programme spanning many orders of magnitude in axion mass and couplings \cite{Irastorza:2018dyq,DiLuzio:2020wdo,Sikivie:2020zpn}.

A large fraction of current axion searches relies on naturally occurring axion populations. Following the original proposal of Sikivie \cite{Sikivie:1983ip}, haloscope experiments search for dark-matter axions in the Galactic halo by converting them into photons in a resonant detector \cite{Adams:2022pbo}, while helioscope experiments look for axions produced in the solar interior and reconverted into X-rays in a laboratory magnetic field \cite{IAXO:2019mpb}. These approaches are extremely powerful and have reached impressive sensitivities. At the same time, their interpretation necessarily depends on assumptions
about the external axion source. Haloscopes require axions to
constitute a sizeable fraction of the local dark-matter density and depend on the phase-space distribution of the Galactic axion field. Helioscopes instead rely on solar production mechanisms and on the modeling of the solar environment. These assumptions are well motivated, but they make the experimental reach partly dependent on astrophysical or cosmological input. For this reason, purely laboratory-based axion searches play a special role because both the production and detection of the new particle occur under controlled conditions, making the interpretation less dependent on assumptions about dark matter or stellar interiors \cite{Jaeckel:2006xm}. This makes laboratory searches particularly valuable as clean probes of the axion-photon interaction, largely independent of astrophysical or cosmological assumptions. Their main limitation is the steep coupling dependence of the signal rate, scaling as $g_{a\gamma}^4$, which reflects the need for both axion production and reconversion. As a result, reaching very small couplings remains challenging even with large magnetic fields, long baselines, and intense photon sources; however, these experiments provide a distinctive and fully laboratory-controlled path to discovery.

The paradigmatic laboratory strategy is the light-shining-through-a-wall (LSW) experiment \cite{Redondo:2010dp}. In its simplest form, photons are converted into axions under a static magnetic field, pass through an opaque barrier, and are then reconverted into photons in a second magnetic field. Since ordinary photons are blocked by the wall, any regenerated light would provide a striking signature of weakly coupled new particles. This idea has led to a variety of LSW implementations and proposals
\cite{Hoogeveen:1990vq,VanBibber:1987rq,Arias:2010bh,Mueller:2009wt}. Optical and near-infrared realisations include ALPS
\cite{Ehret:2010mh,Bahre:2013ywa} and OSQAR \cite{OSQAR:2007oyv}, while related concepts have been explored at microwave and sub-THz frequencies, respectively in CROWS \cite{Betz:2013dza} and STAX \cite{Capparelli:2015mxa}. More recently, the first science results of the ALPS II collaboration have become also available \cite{ALPSII:2025eri}. Other purely laboratory probes are provided by vacuum magnetic birefringence and dichroism experiments, such as PVLAS \cite{PVLAS:2007wzd,Ejlli:2020yhk} and BMV \cite{Cadene:2013bva}, in which axion-like particles would induce polarization changes of light propagating through a strong magnetic field. These searches are complementary to LSW experiments: rather than looking for regenerated photons behind a wall, they test the same axion-photon interaction through precision polarimetry in a controlled laboratory environment. Axion-photon conversion in plasmas has also been considered as a way to exploit tunable effective photon
masses and collective EM modes \cite{Tercas:2018gxv,Mendonca:2019eke,McDonald:2019wou}.

Despite their conceptual cleanliness, standard LSW searches are limited by the small probability of photon-axion-photon conversion, whose improvement typically relies on large magnetic baselines, high photon fluxes, optical cavities, or resonant regeneration. This motivates alternative laboratory
strategies in which the axion signal is generated locally and enhanced by tunable field configurations rather than by the physical size of the apparatus.

High-intensity lasers offer a promising route in this direction. Current and next-generation laser facilities, such as the Extreme Light Infrastructure (ELI) \cite{mourou2011eli}, are approaching regimes in which controlled laboratory studies of strong-field QED and laboratory astrophysics become possible. For instance, the ELI-NP laser facility provides peak powers as high as 10 PW at an operating carrier wavelength of 800 nm with a pulse duration of 24 fs. Considering a focus size of 1 $\mu$m, the peak impinging intensity reaches $10^{24} \; \rm W/cm^2$, implying $E_0\simeq 10^{15}$ V$/$m (electric) and $B_0\simeq 10^{7}$ T (magnetic) field amplitudes. Such strong fields open the possibility of probing nonlinear (NL) quantum vacuum
effects \cite{Marklund:2006my,Fedotov:2006ii,Fedotov:2022ely} and of designing new laser-based searches for weakly coupled particles
\cite{Mendonca:2007zz,Tam:2011kw,Yavuz:2022qbz,Dobrich:2010hi,beyer2020axion,Beyer:2021mzq}. In this context, the large photon occupation numbers, short pulse durations, and tunable geometries of modern laser systems provide new handles with which to engineer axion-photon conversion in vacuum \cite{Dobrich:2010hi,beyer2020axion,Beyer:2021mzq}.
The same NL QED response can also generate third-harmonic photons in non-collinear laser configurations through photon merging
\cite{SundqvistKarbstein2023}, which constitutes an irreducible
Standard-Model contribution to the final state (considered in this work and discussed in Sec.~\ref{sec:sensitivity}).

In this paper, we propose a new experimental design in which axions are actively produced in the laboratory by two intense, polarised, non-collinear laser beams. The central idea is that such an EM field configuration generates a dynamical axion background, depending both on space and time. EM propagation through this light-induced axion background then gives rise to scattered EM oscillations at multiple odd harmonics of the driving frequency, as a consequence of the centrosymmetric nature of the axion-mediated NL interaction, with third-harmonic generation (THG) representing the leading and most efficient process. Although harmonic-generation mechanisms have remained largely unexplored in electrodynamics-based axion searches, they constitute a cornerstone of NL optics. Indeed, second-harmonic generation (SHG) - the leading-order NL process in non-centrosymmetric media - was the first NL optical effect experimentally observed following the advent of the laser \cite{Franken1961}. Starting from the axion-modified NL Maxwell equations, we analytically derive the axion-induced THG field and show that the signal can become detectable for realistic high-intensity laser parameters. A distinctive feature of the proposed setup is that the axion-photon conversion can be resonantly enhanced by tuning the angle between the two laser beams. This angular resonance allows the experiment to scan the axion mass without relying on a resonant detector volume like, e.g., resonant cavities. In this sense, the relevant kinematic matching is controlled by the geometry of the laser configuration rather than by the physical size of the apparatus. 
We find that, for peak laser intensities of the order of
$10^{24}\,{\rm W/cm^2}$, the proposed scheme can probe axion--photon
couplings in the approximate range
$10^{-7}\,{\rm GeV}^{-1}\lesssim |g_{a\gamma}|
\lesssim 10^{-3}\,{\rm GeV}^{-1}$
over the broad mass interval
$m_a\sim10^{-6} \text{-- few}\ {\rm eV}$.

The remainder of this paper is organised as follows. 
After recalling the axion-modified Maxwell equations in 
Sec.~\ref{sec:axionED}, we describe the proposed laser setup configuration and the basic physical mechanisms. In Sec.~\ref{sec:thirdharmonic} we derive the axion-induced THG
field and identify the resonant THG enhancement conditions. In
Sec.~\ref{sec:sensitivity}, we estimate the projected sensitivity in the $(m_{a},g_{a\gamma})$ plane and compare it with existing laboratory and astrophysical constraints.

\section{Axion electrodynamics}

\label{sec:axionED}

\begin{figure*}[t!]
\centering
\begin{center}
\includegraphics[width=\textwidth]{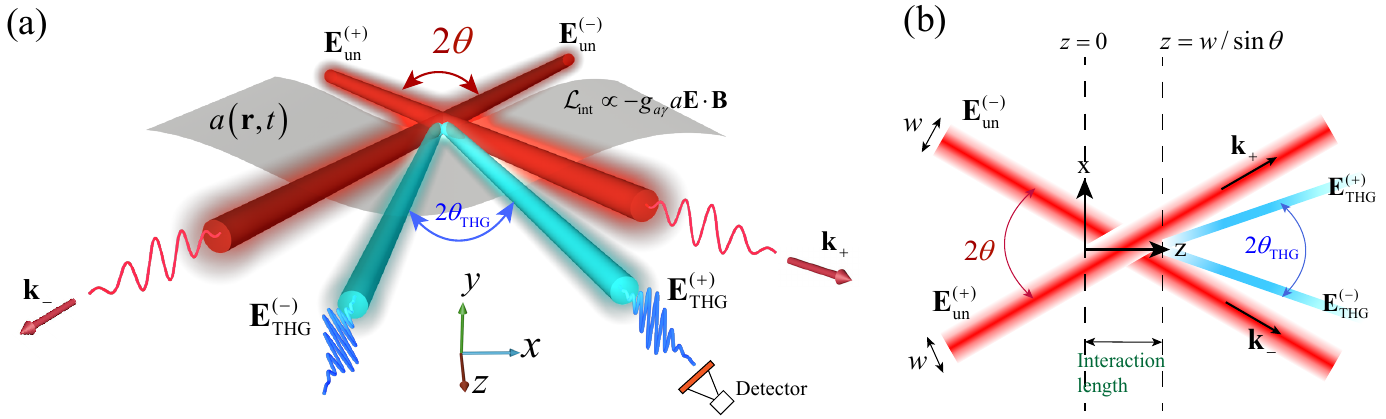}
\caption{\textbf{(a)} Schematic of the proposed configuration. Two non-collinear, high-intensity laser pulses with carrier angular frequency $\omega$ intersect at an angle $2\theta$, with electric-field vectorial amplitudes  $\textbf{E}_{\rm un}^{(\pm)}$, arbitrary polarisations, and wavevectors $\textbf{k}_{\pm}=k_0 \left( \cos\theta \hat{\textbf{z}} \pm \sin \theta \hat{\textbf{x}} \right)$. Their interaction generates a $2\omega$ time-dependent axion field producing two additional THG fields $\textbf{E}_{\rm THG}^{(\pm)}$ scattered at $\pm \theta_{\rm THG} = \pm{\rm tan}^{-1}[{\rm tan}(\theta)/3]$. \textbf{(b)} Side view of the setup, highlighting the interaction length $w/\sin \theta$, where $w$ is the beam width of the incident beam.}		
\end{center}
\label{FIG1}
\end{figure*}

From this point on, we employ SI-normalized electromagnetic and axion
fields. Accordingly, throughout the remainder of the paper, $g_{a\gamma}$ denotes the coupling in the corresponding SI
normalization.
We consider a hypothetical axion field $a(\textbf{r},t)$, which interacts with an EM field through the Lagrangian density
\begin{equation} 
\label{Lagrangian}
\mathcal{L}= \mathcal{L}_{\rm EM} + \mathcal{L}_{a} + \mathcal{L}_{\rm int},
\end{equation}
where $\mathcal{L}_{\rm EM}=-(1/4 \mu_0) F_{\mu \nu} F^{\mu \nu} -  j_e^{\mu}A_\mu$ is the EM Lagrangian, $A^{\mu}$ is the four-vector potential, $F_{\mu \nu}=\partial_\mu A_\nu - \partial_\nu A_\mu$ is the EM field tensor, $j_e^{\mu}$ is the electron four-current, $\mathcal{L}_{a}=(1/2)\big[\partial_\mu a \partial^{\mu}a-(m^2 c^2 a^2/\hbar^2)\big]$ is the free axion Lagrangian density, $m_{a}$ is the axion mass, $c$ is the speed of light in vacuum, $\epsilon_0$ and $\mu_0$ are the vacuum dielectric permittivity and magnetic permeability, respectively, and $\hbar$ is the reduced Planck constant. The Lagrangian density describing the axion-photon interaction is 
\begin{equation}    
\mathcal{L}_{\rm int} = - \frac{1}{4\mu_0}  g_{a\gamma}\,a F^{\mu \nu} \tilde{F}_{\mu \nu}, 
\end{equation} 
where $\tilde{F}_{\mu \nu}=(1/2)\varepsilon_{\mu \nu \rho \sigma}F^{\rho \sigma}$ ($\varepsilon_{0123}=+1$ is the Levi-Civita tensor) is the dual EM tensor. Applying the Euler-Lagrange equations, one obtains the modified Maxwell's equations (MMEs) in SI units
\begin{subequations} 
\label{MEaxion} 
\begin{align} 
& \nabla \cdot \textbf{E} =  \; \frac{\rho}{\epsilon_0} -  c g_{a\gamma} \textbf{B} \cdot \nabla a , \\
& \nabla \cdot \textbf{B}  = \; 0 , \\
& \nabla \times \textbf{E} +  \dot{\textbf{B}} = \; 0 , \\
& \nabla \times \textbf{B} - \frac{1}{c^2}  \dot{\textbf{E}} = \; \mu_0 \textbf{J} + \frac{g_{a\gamma}}{c} \textbf{B} \dot{a} - \frac{g_{a\gamma}}{c} \textbf{E} \times \nabla a ,
\end{align}
\end{subequations}
where the dot denotes the time derivative $\partial_t$, $c=1/\sqrt{\epsilon_0 \mu_0}$ is the speed of light in vacuum, $\rho$ and $\textbf{J}$ are the charge and current density, respectively. Here, we consider ideal ``vacuum" conditions where $\rho=0$ and $\textbf{J}=0$. However, the presence of the axion field produces additional charge and current densities, which  modify Maxwell's equations nonlinearly. From the Euler-Lagrange equations, one also obtains an inhomogeneous Klein-Gordon equation for the axion field  
\begin{equation}
\label{KGphi}
\left(\square +  \frac{m_{a}^2 c^2}{\hbar^2}\right) a = \frac{g_{a \gamma}}{\mu_0 c} \textbf{E} \cdot \textbf{B} \, ,
\end{equation}
where the inhomogeneous term involves the scalar product of  electric ($\textbf{E}$) and magnetic ($\textbf{B}$) fields and $\square \equiv \partial^{\mu}\partial_{\mu} = c^{-2}\partial_t^2-\nabla^2$ indicates the d'Alembert operator. Note that, if $\textbf{E},\textbf{B}$ are quasi-monochromatic fields with carrier angular frequency $\omega$, the $\textbf{E} \cdot \textbf{B}$ axion generation term produces both quasi-static and $2\omega$ time-oscillating axion fields, which play a central role in our proposed setup exploiting axion-induced THG, see below.

\section{Third-harmonic generation}
\label{sec:thirdharmonic}

Figs.~1(a,b) illustrate the concept of the proposed experimental setup.  We consider two high-intensity laser pulses with electric field vectorial envelopes $\textbf{E}_0^{(\pm)}({\bf r},t)$ making an angle $2\theta$ between them. For the $\simeq 24$ fs pulses available at the ELI-NP laser facility, the spectral bandwidth $\Delta\omega$ remains sufficiently narrow compared with the carrier angular frequency $\omega$ to justify a quasi-monochromatic approximation $\Delta\omega \ll \omega$. Moreover, since propagation occurs in vacuum, material dispersion does not introduce any additional spectral reshaping of the pulses. Although focusing to spot sizes as small as $w \simeq 1~\mu{\rm m}$ would in principle lead to appreciable diffraction of the two non-collinear pulses, axion generation is confined to their spatial overlap region, whose longitudinal extent is approximately $w/\sin\theta$ [see Figs.~1(a,b)], where ${\bf E}\cdot{\bf B}$ can be non-vanishing due to the non-collinear excitation. Over this limited interaction length, diffraction can therefore be neglected, allowing the two pulses to be modelled as plane waves. In turn, the unperturbed electric field $\textbf{E}_{\rm un}(\textbf{r},t)$ accounting for the coherent superposition of the two impinging laser beams is given by
\begin{align}
\label{E0}
\textbf{E}_{\rm un} = \text{Re}\left\{ \Big[ E_+ \hat{\bf n}_+ e^{i\textbf{k}_+ \cdot \textbf{r} } + E_-\hat{\bf n}_- e^{i\textbf{k}_- \cdot \textbf{r} }   \Big] e^{-i\omega t} \right\},
\end{align} 
where the $\pm$ index labels the two distinct laser fields with $E_{\pm}$ electric field vectorial amplitudes, polarisation unit vectors $\hat{\bf n}_{\pm} = \alpha_{\rm TM}^{(\pm)}\cos \theta \hat{\textbf{x}} +  \alpha_{\rm TE}^{(\pm)} \hat{\textbf{y}} \mp \alpha_{\rm  TM}^{(\pm)} \sin \theta \hat{\textbf{z}}$, and wavevectors $\textbf{k}_{\pm} = k_z \hat{\textbf{z}} \pm k_x  \hat{\textbf{x}}$, where $k_z = k_0\cos\theta $, $k_x = k_0\sin \theta$, $k_0=\omega/c$ is the vacuum wavenumber, and $\hat{\bf x},\hat{\bf y},\hat{\bf z}$ indicate the $x,y,z$ unit vectors. The complex coefficients $\alpha_{\rm TE}^{(\pm)}$ and $\alpha_{\rm TM}^{(\pm)}$ (such that $|\alpha_{\rm TE}^{(\pm)}|^2 + |\alpha_{\rm TM}^{(\pm)}|^2 = 1$) account for the arbitrary polarisation state of the two impinging beams. The corresponding unperturbed magnetic field is calculated as $\textbf{B}_{\rm un}(\textbf{r},t)=\text{Re}\{\omega^{-1} \Big[ E_+ \textbf{k}_+\times \hat{\bf n}_+ e^{i\textbf{k}_+ \cdot \textbf{r} } + E_-\textbf{k}_-\times\hat{\bf n}_- e^{i\textbf{k}_- \cdot \textbf{r} }   \Big] e^{-i\omega t}\}$. Owing to the weak axion-photon coupling $g_{a\gamma}$, at first order in perturbation theory in the $g_{a\gamma}$ parameter, the unperturbed impinging field produces a weak axion field $a({\bf r},t)\propto g_{a\gamma}$ satisfying 
\begin{equation}
\label{pertaxioneq}
\left(\square +  \frac{m_{a}^2 c^2}{\hbar^2}\right) a = \frac{g_{a \gamma}}{\mu_0 c} \textbf{E}_{\rm un} \cdot \textbf{B}_{\rm un}.
\end{equation}
The solution of such an inhomogeneous partial differential equation (PDE) is provided by the superposition of particular solutions $\propto e^{2ik_z z -2i\omega t},e^{2ik_x x}$ and forward/backward homogeneous solutions $\propto e^{\pm 2ik_0 \gamma_{a} z - 2i\omega t}, e^{2ik_x x \mp 2 k_0 z \sqrt{\mu_{a}^2 - \cos^2\theta} }$ (where $\gamma_{a} = \sqrt{1-m_{a}^2c^4/4\hbar^2\omega^2}$ and $\mu_{a} = \sqrt{1+m_{a}^2c^4/4\hbar^2\omega^2}$) with arbitrary amplitudes, to be determined by boundary conditions (BCs). Because such amplitudes depend on the longitudinal wavenumber components [$z$-components, see 
Figs.~1(a,b)] mismatch, one can safely neglect the generation of backward axions (producing large longitudinal wavenumber mismatch with respect to the longitudinal component $2k_z$ of the $2\omega$ contribution to ${\bf E} \cdot {\bf B}$) and evaluate only forward axion generation through the sole $a({\bf r}_\bot,z=0,t) = 0$ BC [beginning of the interaction region, see 
Figs.~1(a,b)], where ${\bf r}_\bot = x\hat{\bf x} + y\hat{\bf y}$, providing  
\begin{widetext}
\begin{eqnarray} \label{phisol}
a(\textbf{r},t)&=&\frac{\hbar^2 g_{a\gamma}}{\mu_0 m_{a}^2 c^4} \text{Re}\left[E_+ E_-^* \left( \alpha_{\rm TE}^{(-)*} \alpha_{\rm TM}^{(+)} + \alpha_{\rm TE}^{(+)}\alpha_{\rm TM}^{(-)*}  \right)\frac{\sin^2 \theta }{1+4\frac{\hbar^2 \omega^2}{m_{a}^2 c^4} \sin^2 \theta  }    \left(1- e^{-2k_0 z \sqrt{\mu_{a}^2 - \cos^2\theta  } } \right) e^{2ik_x x } \right] + \nonumber \\   &+& \frac{\hbar^2 g_{a\gamma}}{\mu_0 m_{a}^2 c^4} \text{Re}\left[ E_+ E_- \left( \alpha_{\rm TE}^{(+)} \alpha_{\rm TM}^{(-)} + \alpha_{\rm TE}^{(-)}\alpha_{\rm TM}^{(+)}  \right) \frac{ \sin^2 \theta }{1-4\frac{\hbar^2 \omega^2}{m_{a}^2 c^4} \sin^2 \theta  }  \left( e^{2ik_z z  }- e^{2ik_0 \gamma_{a}z }    \right) e^{- 2i\omega t} \right],  
\end{eqnarray}
\end{widetext}
see full derivation in Appendix \ref{appA}. Note that Eq.~(\ref{phisol}) contains both a static part and a dynamic contribution oscillating with angular frequency $2\omega$ (i.e., the term $\propto e^{-2i\omega t}$), which is the one responsible for THG, see below. Note also in Eq. (\ref{phisol}) that the $e^{-2i\omega t}$ coefficient can {\it resonate} for $\gamma_{a} = \cos \theta$, occurring when  
\begin{equation}
    \label{rescond}
    \sin \theta = \frac{m_{a} c^2}{2 \hbar \omega},
\end{equation}
i.e., when longitudinal phase-matching occurs between the $2\omega$ components of $a({\bf r},t)$ and ${\bf E}\cdot{\bf B}$. Such a resonance, peculiar of the non-collinear laser field excitation conditions illustrated in 
Figs.~1(a,b), happens when the phase of axions generated in distinct points of space matches the one of the ${\bf E}\cdot{\bf B}$ driving term, leading to progressive axion field build-up rather than successive constructing/destructing interferences, similarly to a harmonic oscillator coherently driven at resonance. Indeed, at the $\gamma_{a} = \cos \theta$ resonance
\begin{equation}
\frac{ e^{2ik_z z}- e^{2ik_0 \gamma_{a}z } }{1-4\frac{\hbar^2 \omega^2}{m_{a}^2 c^4} \sin^2 \theta  } \rightarrow  - i\sin \theta {\rm tan}\theta k_0 z e^{2ik_0 {\rm cos}\theta  z},
\end{equation}
and the $2\omega t$ (temporal), $2k_0 {\rm cos}\theta z$ (spatial) axion oscillations grow linearly with $z$. The apparent divergence as $\theta\rightarrow\pi/2$ is an artifact of the monochromatic plane-wave approximation and does not correspond to a physical enhancement. In this limit, the nominal interaction length $w/{\rm sin}\theta$ no longer provides an adequate description of the actual spatio-temporal overlap between the two colliding laser fields: in this counter-propagating regime the interaction length becomes large, and thus the neglect of diffraction, pulse overlap (the maximum overlap in counter-propagating regime is $\simeq c t_0 \simeq 10$ $\mu$m, where $t_0 \simeq 24$ fs is the pulse duration) and quasi-monochromatic treatment breaks down.

At second order in perturbation theory over the $g_{a\gamma}$ parameter, the axion field $a({\bf r},t)\propto g_{a\gamma}$ produces a perturbed EM field $\textbf{E}_{\rm pert}(\textbf{r},t)\propto g_{a\gamma}^2$, explicitly given by
\begin{align}
\label{ET}
\textbf{E}_{\rm pert}(\textbf{r},t) &= \text{Re}\Big[  \textbf{E}_{\rm THG}(\textbf{r})e^{-3i\omega t} 
+ \textbf{E}_{\rm Kerr}(\textbf{r})e^{-i\omega t} \Big],
\end{align}
i.e., a superposition of $\omega$ (the so-called Kerr effect) and $3\omega$ (THG) time-oscillating NL contributions with vectorial spatial profiles $\textbf{E}_{\rm Kerr}({\bf r})$ and $\textbf{E}_{\rm THG}({\bf r})$, respectively. While the Kerr effect produces wave-mixing worth of future investigation, in this article we focus solely on THG, and thus in what follows we ignore the Kerr field happening at the distinct angular frequency $\omega$. Physically, the THG signal can be understood as arising from the propagation of the fundamental optical field through a space- and time-dependent axion background oscillating at $2\omega$, which effectively modulates the EM response and generates a component at $3\omega$.  

Owing to the resonant photon-axion interaction in the considered non-collinear configuration, see Eqs.~(\ref{phisol},\ref{rescond}), in what follows we fix everywhere the resonance condition yielding the maximum number of THG ($3\omega$) photons for given excitation angle (dependent on the axion mass $m_{a}$, and thus to be scanned in order to identify $m_{a}$) and polarisation. At $g_{a\gamma}^2$ order, Eqs.~\ref{MEaxion} provide 
\begin{subequations}
\begin{flalign} 
\label{ebres}
\textbf{E}_{\rm THG}^{\rm (res)}(\textbf{r}) & = \frac{c}{v_0} \left[E_+ {\bf p}_{\rm e}^{+}(z)e^{ik_x x} + E_- {\bf p}_{\rm e}^{-}(z)e^{-ik_x  x} \right], &&\raisetag{0.8\baselineskip}\\
\textbf{B}_{\rm THG}^{\rm (res)}(\textbf{r}) & = \frac{1}{v_0} E_+ {\bf p}_{\rm b}^{+}(z)e^{ik_x x}  + \frac{1}{v_0} E_- {\bf p}_{\rm b}^{-}(z)e^{-ik_x x} , &&\raisetag{0.8\baselineskip}
\end{flalign}
\end{subequations}
where 
\begin{subequations}
\begin{align}
{\bf p}_{\rm e}^{\pm}(z) & = \alpha_{\rm TE}^{(\pm)} \mathcal{A}_x(z) \hat{\textbf{x}}  + \alpha_{\rm TM}^{(\pm)} \mathcal{A}_y(z)  \hat{\textbf{y}}  \pm  \alpha_{\rm TE}^{(\pm)} \mathcal{A}_z(z)  \hat{\textbf{z}}, &&\raisetag{0.9\baselineskip}\\ 
{\bf p}_{\rm b}^{\pm}(z) & = \alpha_{\rm TE}^{(\pm)} \mathcal{B}_x(z) \hat{\textbf{x}}  + \alpha_{\rm TM}^{(\pm)} \mathcal{B}_y(z)  \hat{\textbf{y}}  \pm  \alpha_{\rm TE}^{(\pm)} \mathcal{B}_z(z)  \hat{\textbf{z}},&&\raisetag{0.9\baselineskip} \\
\mathcal A_x(z) & = 3\left[\Gamma_{a} e^{ik_0\Gamma_{a}z} - (\eta_{a}-iZ) e^{3i\gamma_{a} k_0 z} \right],
\label{coeffs:Ax} \\
\mathcal A_y(z) & = 9\left[ \frac{\eta_{a}}{\Gamma_{a}} e^{ik_0\Gamma_{a}z} - \left( 1 + i\frac{Z}{\gamma_{a}}
\right) e^{3i \gamma_{a} k_0z} \right], \label{coeffs:Ay} \\
\mathcal A_z(z) & = 3\frac{k_x}{k_0}
\left[ \left( 1 - 3i\frac{Z}{\gamma_{a}} \right) e^{3ik_0 \gamma_{a} z} - e^{ik_0\Gamma_{a}z} \right], \label{coeffs:Az} \\
\mathcal B_x(z) & = 3 \left[ (\eta_{a} + 3i Z ) e^{3i k_0 \gamma_{a} z} - \eta_{a} e^{ik_0\Gamma_{a}z} \right], \label{coeffs:Bx} \\
\mathcal B_y(z) & = 3\left[ 3 e^{ik_0\Gamma_{a}z} - \left( 
3 - i\frac{Z}{\gamma_{a}} \right) e^{3i k_0 \gamma_{a} z} \right], \label{coeffs:By} \\
\mathcal B_z(z) & = 3 \frac{k_x}{k_0} \left[ \frac{\eta_{a}}{\Gamma_{a}} e^{ik_0\Gamma_{a}z} - \left( 1+i \frac{Z}{\gamma_{a}} \right) e^{3i \gamma_{a} k_0z} \right], \label{coeffs:Bz}
\end{align}
\end{subequations}
$v_0^{-1}=\frac{g_{a\gamma}^2 E_+E_-}{384c\mu_0 \omega^2} [\alpha_{\rm TE}^{(-)} \alpha_{\rm TM}^{(+)} + \alpha_{\rm TE}^{(+)} \alpha_{\rm TM}^{(-)}]$, $Z= 4 k_0 z (1-\gamma_{a}^2)$, $\Gamma_{a} = \sqrt{8+\gamma_{a}^2}$, and $\eta_{a} = (4-\gamma_{a}^2)/\gamma_{a}$, see the Appendix \ref{appB} for the full derivation. 
These expressions provide the forward-propagating solution for the
electric and magnetic THG fields at order $g_{a\gamma}^{2}$ and at the
resonance condition, within the monochromatic plane-wave and
unidirectional approximations employed here. 
Note that, also in the THG electric and magnetic fields, there are resonant contributions (growing with $z$) originated by the resonant photon-axion interaction. In turn, at resonance, the ${\cal A}_{x,y,z}(z)$ and ${\cal B}_{x,y,z}(z)$ spatial oscillations are dominated by $e^{3i \gamma_{a} k_0z}$ contributions that are the only ones growing linearly with $z$, and the THG-EM field is practically scattered at the angles $\pm \theta_{\rm THG}$, where $ \theta_{\rm THG} = {\rm tan}^{-1}[{\rm tan}(\theta)/3]$. Such an angular separation is highly important for our proposed axion detection scheme, as it produces the spatial separation of THG and driving field spots, thus enabling long-time THG acquisition.

From a practical detection perspective, one is interested in the number of THG photons emitted at resonance. Assuming Gaussian  intensity profiles (both in space and time) of the impinging driving fields, the number of THG emitted photons per shot at $\theta_{\rm THG}$ is 
\begin{equation} 
N_{\rm THG}^{\rm (res)}=\frac{1}{3\hbar \omega} \iint d^2\textbf{r}  dt \; I_{\rm THG}^{\rm (res)}(\textbf{r},t),
\label{nphoton} 
\end{equation}
where $I_{\rm THG}^{\rm (res)}(\textbf{r},t)= (c/\mu_0|v_0|^2) |E_+|^2 \rm \lvert Re\left[  {\bf p}_{\rm e}^{+}\times {\bf p}_{\rm b}^{+,*}\right]\rvert$ is the THG intensity at $\theta_{\rm THG}$ (the intensity at $-\theta_{\rm THG}$ coincides with $I_{\rm THG}^{\rm (res)}$ due to the exchange symmetry of the non-collinear pump fields). By retaining only the resonant terms ($\propto z$)  in the THG fields above, we obtain an approximate expression of $N_{\rm THG}^{\rm (res)}$ for cross polarised $[\alpha_{\rm TE}^{(+)}=1, \alpha_{\rm TM}^{(+)}=0,\alpha_{\rm TE}^{(-)}=0, \alpha_{\rm TM}^{(-)}=1]$ Gaussian beams/pulses after propagating through the interaction distance $z=w/\sin \theta$, explicitly
\begin{equation} 
N_{\rm THG}^{\rm (res)} \simeq \frac{1}{3} \sqrt{\frac{2{\rm ln}2}{\pi}} \left(\frac{1}{16\pi^2}\right)^2\sin \theta \tan \theta\frac{g_{a \gamma}^4}{\hbar c^3}    \frac{P_0^3 \lambda^3 t_0}{w^2},
\label{nthgapprox} 
\end{equation}
where $\lambda=2 \pi c/\omega$ is the carrier wavelength of the driving pulses, $w$ is the full width at half maximum (FWHM) of the spatial spot, $P_0$ is the peak power, and $t_0$ is the FWHM duration of the impinging pulses.  
Here and in the following,
$N_{\rm THG}^{(\rm res)}$ 
denotes only the axion-mediated
contribution to the $3\omega$ photon yield. The Euler--Heisenberg
interaction can generate an additional coherent $3\omega$ field.
Consequently, the total detected intensity is proportional
to
%
  $\left|
    \mathcal{A}_{a}+\mathcal{A}_{\rm EH}
  \right|^{2}$,
%
and not to an incoherent sum of the axion and QED photon numbers.
The treatment of this Standard-Model contribution is discussed in
Sec.~\ref{sec:sensitivity}.

\begin{figure}[t!]
\centering
\begin{center}
\includegraphics[width=\columnwidth]{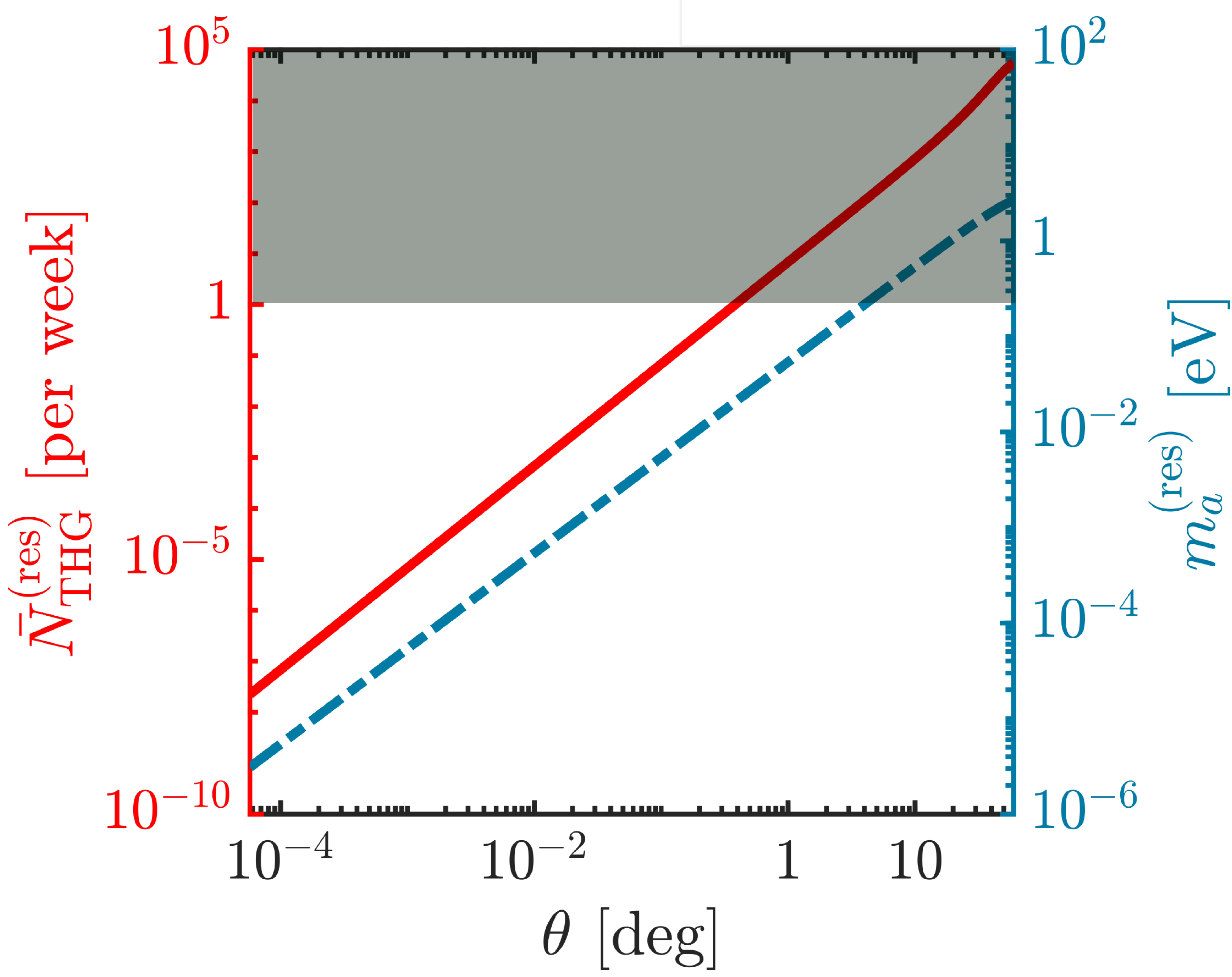}
\caption{Non-collinear pump angle $\theta$ dependence of the time-integrated (observation time $\Delta t=1$ week) THG photon number $\bar{N}_{\rm THG}^{(\rm res)}$ (red solid line, left vertical axis) emitted at $\theta_{\rm THG}={\rm tan}^{-1}[{\rm tan}(\theta)/3]$ in resonance conditions $m_{a}^{\rm (res)} = (2\hbar \omega/c^2) \sin  \theta$ (blue dashed line, right vertical axis), after propagating through the interaction length $z=w/\sin \theta$. The impinging pulse characteristics are the ones of ELI-NP laser focused pulsed beams with carrier wavelength $\lambda=800$ nm, beam waist (FWHM) $w=1 \; \mu \rm m$, FWHM pulse duration $t_0=24 \; \rm fs$, repetition rate $\Gamma=$ 1 shot per minute, peak power $P_0=10$ PW, and cross-linear polarised impinging fields with polarisation coefficients $\alpha_{\rm TE}^{(+)}=1, \alpha_{\rm TM}^{(+)}=0,\alpha_{\rm TE}^{(-)}=0, \alpha_{\rm TM}^{(-)}=1$. The axion-photon coupling constant considered is $g_{a\gamma}=10^{-5} \; \rm GeV^{-1}$. The shaded part indicates the single-photon-yield region $\bar{N}_{\rm THG}^{(\rm res)} \ge 1$. }
\label{fig2}
\end{center}
\end{figure}

\begin{figure}[hbt!]
\centering
\begin{center}
\includegraphics[width=\columnwidth]{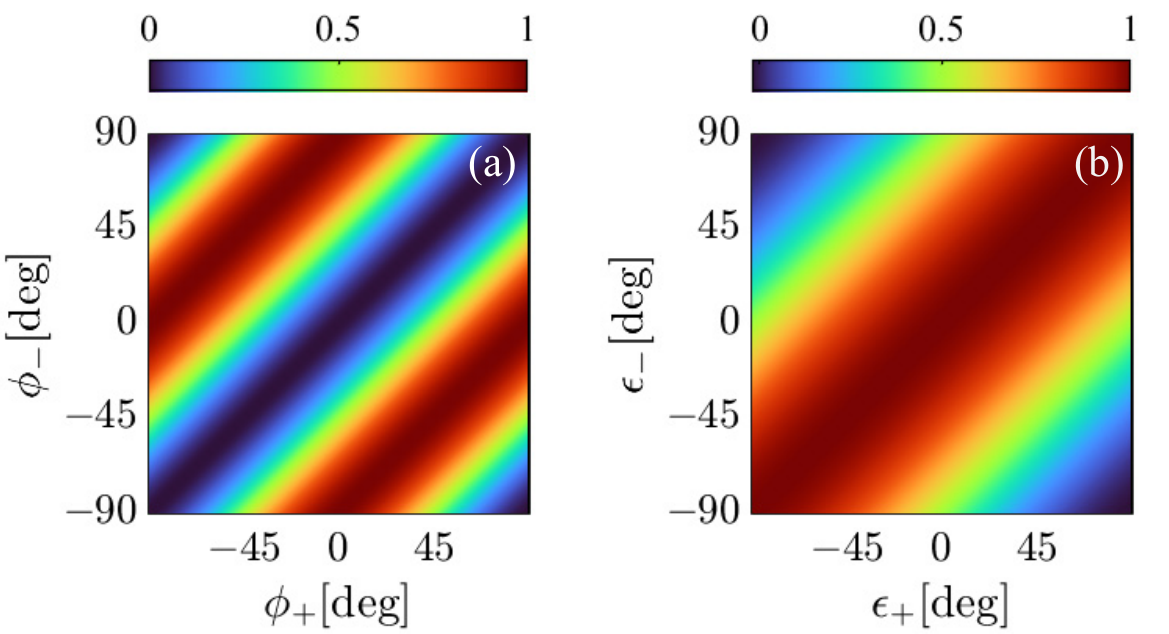}
\caption{ Dependence of the time-integrated THG photon number  $\bar{N}_{\rm THG}^{(\rm res)}$, emitted at resonance for  $\theta=$ 10° and $z=w/\sin \theta$, on \textbf{(a)} the linear polarisation angles $\phi_+, \phi_-$  $(\rm with \; \alpha_{\rm TE}^{(+)}=\cos \phi_+, \, \alpha_{\rm TM}^{(+)}=\sin \phi_+, \, \alpha_{\rm TE}^{(-)}=\cos \phi_-, \, \alpha_{\rm TM}^{(-)}=-\sin \phi_-)$. The maxima of $\bar{N}_{\rm THG}^{(\rm res)}$ are observed for $|\phi_+-\phi_-|=\pi/2$. \textbf{(b)} $\bar{N}_{\rm THG}^{(\rm res)}$ dependence on the ellipticities $\epsilon_+, \epsilon_-$, where $\alpha_{\rm TE}^{(+)}=e^{i\epsilon_+}/\sqrt{2}, \alpha_{\rm TM}^{(+)}=1/\sqrt{2},\alpha_{\rm TE}^{(-)}=e^{i\epsilon_-}/\sqrt{2}, \alpha_{\rm TM}^{(-)}=1/\sqrt{2}$ with maxima at $\epsilon_+=\epsilon_-$. The colorbars are normalized by the maximum value.  The parameters used in the simulation refer to ELI-NP pulse characteristics, i.e., carrier wavelength $\lambda=800$ nm, beam waist (FWHM) $w=1 \; \mu \rm m$, pulse FWHM duration $t_0=24 \; \rm fs$,  repetition rate $\Gamma=$ 1 shot per minute,  peak power $P_0=10$ PW, and 1 week of observation time. The axion-photon coupling constant considered in the plots is $g_{a\gamma}=10^{-5} \; \rm GeV^{-1}$. }
		\label{fig3}
	\end{center}
\end{figure}

\begin{figure}[hbt!]
\centering
\begin{center}
\includegraphics[width=\columnwidth]{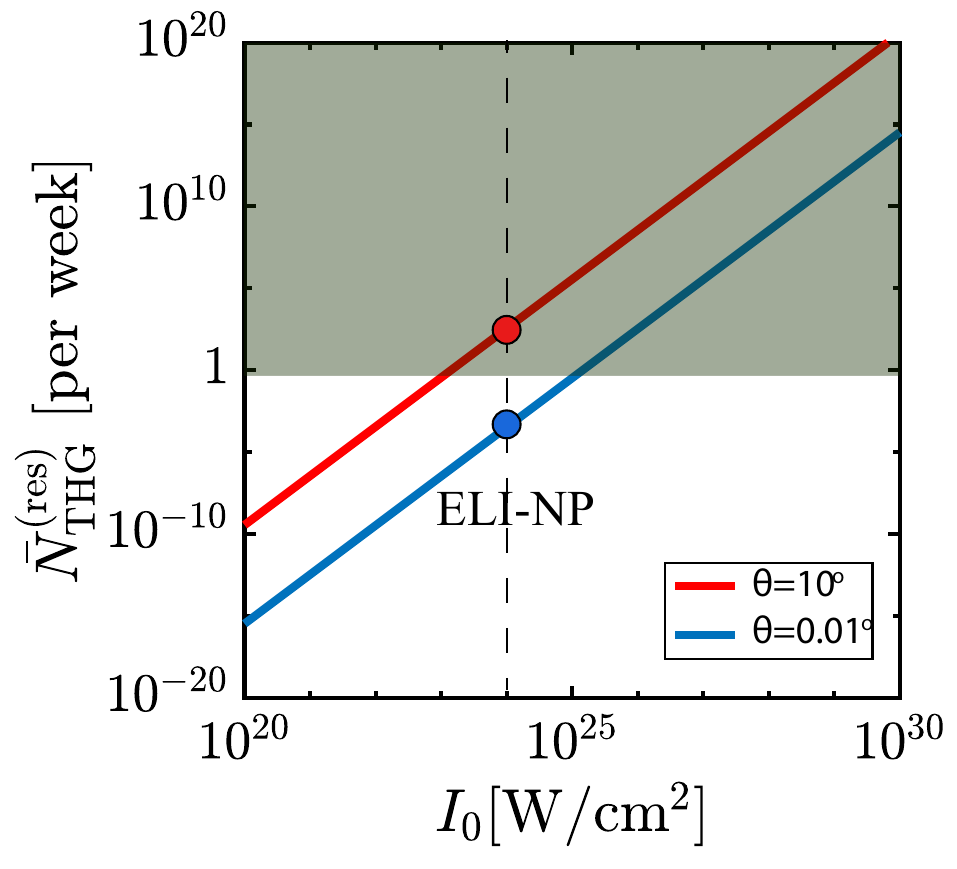}
\caption{Incident intensity $I_0$ dependence of integrated number of third harmonic photons $\bar{N}_{\rm THG}^{(\rm res)}$  emitted at resonance for two distinct values of the beam angle $\theta=0.01^{\circ}$ (blue), and $\theta=10^{\circ}$ (red), calculated at a distance $z=w/\sin \theta$. Considered center wavelength is $\lambda=800$ nm, beam waist $w=1 \; \mu \rm m$, temporal pulse width $t_0=24 \; \rm fs$, pulse repetition rate $\Gamma =1/60$ Hz, observation time $\Delta t=1$ week,  and cross-linear polarised impinging beams with polarisation coefficients $\alpha_{\rm TE}^{(+)}=1, \alpha_{\rm TM}^{(+)}=0,\alpha_{\rm TE}^{(-)}=0, \alpha_{\rm TM}^{(-)}=1$. The shaded region represents the single-photon-yield region ($\bar{N}_{\rm THG}^{(\rm res)} \ge 1 $). The dashed vertical line represents the peak intensity of $I_0 \approx 10^{24} \; \rm W/cm^2 $ corresponding to the ELI-NP peak power of $P_0=10$ PW.  }
		\label{fig4}
	\end{center}
\end{figure}

For numerical evaluation, the SI-normalized coupling is related to the
conventional coupling $g_{a\gamma}^{\rm conv}$, quoted in
${\rm GeV}^{-1}$ and labelled simply $g_{a\gamma}$ in the sensitivity
plots, by
\begin{equation}
g_{a\gamma}^{\rm SI}
=
\sqrt{\hbar c}\,
\left(
\frac{g_{a\gamma}^{\rm conv}}
     {1\,{\rm GeV}^{-1}}
\right)
\frac{1}{1.602176634\times10^{-10}\,{\rm J}} .
\end{equation}
Thus, the coupling appearing in Eq.~(15) is
$g_{a\gamma}^{\rm SI}$, with dimensions $({\rm m/J})^{1/2}$; numerical
values quoted in ${\rm GeV}^{-1}$ are converted using the relation above
before being inserted into Eq.~(15).

In the considered axion detection scheme, the quantity of greater importance is the integrated photon number  $\bar{N}_{\rm THG}^{(\rm res)}=N_{\rm THG}^{(\rm res)} \Gamma \Delta t $, where $\Delta t$ is the observation time and $\Gamma$ is the pulse repetition rate. Fig.~\ref{fig2} depicts the excitation angle $\theta$ dependence of the THG photon number that can be obtained from two cross-polarised ELI-NP beams operating at peak power $P_0=10 $ PW, observation time of $\Delta t=1$ week, and pulse repetition rate of $\Gamma=1/60$ Hz, i.e., 1 shot per minute. As predicted by 
Eq.~(\ref{nthgapprox}), we observe an increase of the emitted THG photon number with the excitation angle $\theta$. Note that the shaded region in the $\bar{N}_{\rm THG}^{\rm (res)}-\theta$ plane in Fig.~\ref{fig2} for which $\bar{N}_{\rm THG}^{\rm (res)} \ge 1$ represents the detectable range anticipated by our proposed experiment. 

Another distinguishable feature of the THG signal is its dependence on the state of polarisation of the impinging laser fields determined by the polarisation coefficients $\alpha_{\rm TE}^{(\pm)}$ and $\alpha_{\rm TM}^{(\pm)}$. For fixed excitation angle $\theta$, the emitted THG photon number $\bar{N}_{\rm THG}^{\rm (res)}$  at resonance achieves maxima for a certain combination of the polarisation coefficients, as depicted in Fig.~\ref{fig3}. In Fig.~\ref{fig3}(a), we depict the dependence of $\bar{N}_{\rm THG}^{\rm (res)}$ on the linear polarisation angles $\phi_{\pm}$, where $ \alpha_{\rm TE}^{(\pm)}=\cos \phi_{\pm}, \, \alpha_{\rm TM}^{(\pm)}=\pm\sin \phi_{\pm}$. Similarly,  Fig.~\ref{fig3}(b) depicts the dependence of  $\bar{N}_{\rm THG}^{\rm (res)}$ on the ellipticities $\epsilon_{\pm}$, where the polarisation coefficients are  $\alpha_{\rm TE}^{(\pm)}=e^{i\epsilon_{\pm}}/\sqrt{2}, \alpha_{\rm TM}^{(\pm)}=1/\sqrt{2}$. For the linear polarisation case, we observe the maxima of $\bar{N}_{\rm THG}^{\rm (res)}$ occurs at $|\phi_+-\phi_-|=\pi/2$, when the beams are cross-polarised  [see Fig.~\ref{fig3} (a)], while for elliptical polarisation, the maximum is observed for $\epsilon_+=\epsilon_-$ [see Fig.~\ref{fig3} (b)]. This behaviour can be traced back to the axion source term ${\bf E}\cdot{\bf B}$, whose amplitude is maximised for cross-polarised excitation, thereby maximising the generated axion field and, consequently, the THG signal.

In addition, it is worth investigating the detectable response of the proposed setup over impinging intensity $I_0=(1/2)\epsilon_0 c \lvert \textbf{E}_0 \rvert^2$, as depicted in Fig.~\ref{fig4} for two distinct beam angles $\theta=10^{\circ}$ and $\theta=0.01^{\circ}$. We observe that $\bar{N}_{\rm THG}^{\rm (res)}$ shows a cubic dependence on $I_0$, as is evident from Eq. (\ref{nthgapprox}). 

\begin{figure}[t!]
\centering
\begin{center}
\includegraphics[width=\columnwidth]{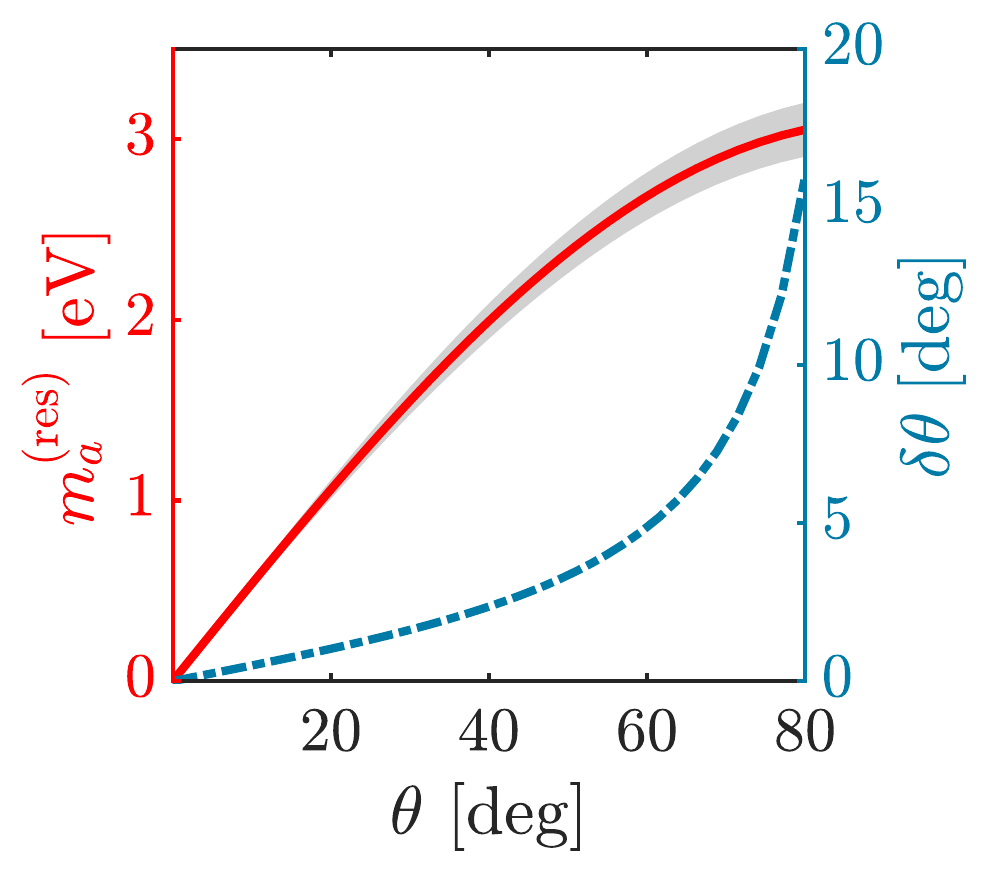}
\caption{Excitation angle $\theta$ dependence of the resonant axion mass $m_{a}^{\rm (res)} = (2\hbar \omega/c^2) \sin  \theta$  (red solid) and the corresponding angular scan step size $\delta \theta$ (blue dashed) defined in Eq. (\ref{scanrate}) to probe the mass region. The parameters used refer to ELI-NP laser beams with a carrier wavelength of $\lambda=800$ nm and spectral width of $\Delta \lambda = 56$ nm. The shaded region indicates the width $\pm \delta m_{a}$. Assuming a minimum angular stepsize $\theta \gtrsim 1 \; \mu\text{rad}$ , the full mass range of $m_{a} \sim 3\times10^{-6}-3.09$ eV can be scanned in approximately 197 steps.}
\label{fig5}
\end{center}
\end{figure}

\begin{figure*}[hbt!]
\centering
\begin{center}
\includegraphics[width=1.8\columnwidth]{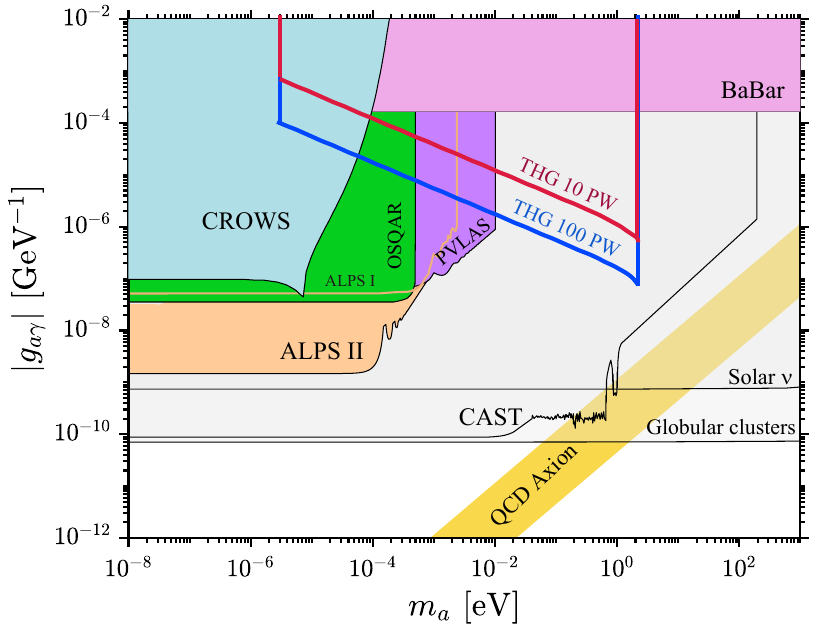}
\caption{Projected single-photon reach of our proposed experiment in the $g_{a\gamma}$ vs $m_{a}$ plane (in red solid line), calculated for cross-polarised ELI-NP beam parameters: peak power 10 PW, carrier wavelength $\lambda=800$ nm, beam waist (FWHM) $w=1 \; \mu \rm m$, FWHM pulse duration $t_0=24 \; \rm fs$, repetition rate $\Gamma=1/60$ Hz, observation time $\Delta t=$ 1 week, and cross-linear polarised impinging beams with  polarisation coefficients $\alpha_{\rm TE}^{(+)}=1, \alpha_{\rm TM}^{(+)}=0,\alpha_{\rm TE}^{(-)}=0, \alpha_{\rm TM}^{(-)}=1$. 
The quoted one-week exposure applies separately to each point of this
point-by-point sensitivity projection.
The blue solid line represents the bounds achievable with a future ELI-like setup with peak power 100 PW, carrier wavelength  $\lambda =1050$ nm, FWHM pulse duration $t_0=30 \; \rm fs$, and repetition rate $1/60$ Hz. Both projections are defined by the idealised single-photon condition
$\epsilon_{\rm det}\bar N_{\rm THG}^{(\rm res)}=1$ and should not be
interpreted as confidence-level exclusions. They assume that the
effective background can be reduced below one event by operating near
an Euler--Heisenberg coherence minimum and by using angular, waist,
spectral, and polarisation control measurements. The residual
finite-pulse QED contribution and its interference with the axion
amplitude are not included in the numerical curves.
The light blue region shows existing bounds from the CROWS experiment \cite{Betz:2013dza}, the green region is excluded by the OSQAR experiment \cite{OSQAR:2007oyv}, the purple region is excluded by PVLAS \cite{PVLAS:2007wzd,Ejlli:2020yhk}, and the pink region is excluded by BaBar \cite{Dolan:2017osp}. The orange solid line represents the ALPS I bounds \cite{Ehret:2010mh,Bahre:2013ywa}, the light orange region represents the constraints from ALPS II \cite{ALPSII:2025eri}. CAST \cite{CAST:2024eil}, 
solar neutrinos \cite{Vinyoles:2015aba} and horizontal branch 
\cite{Ayala:2014pea} rely on astrophysical sources and are represented by the light grey regions. 
The darker yellow band represents the QCD axion region, following the definition in \cite{DiLuzio:2016sbl,DiLuzio:2017pfr}.
}
\label{fig6}
\end{center}
\end{figure*}

\section{Projected bounds sensitivity}
\label{sec:sensitivity}
In this section, we elucidate the anticipated reach of the above proposal adopting the ELI-NP laser parameters. As emphasized in the previous sections, the proposed setup is most sensitive at the resonance condition  $m_{a}^{\rm (res)} = (2\hbar \omega/c^2) \sin  \theta$, see Eq.~(\ref{rescond}), which depends on the carrier frequency $\omega$ of the impinging laser beams.  In real laser systems, the finite spectral width $\Delta \omega$ ensures axion mass detuned from the carrier frequency are produced at a decreasing rate. Consequently, each value of $\theta$ corresponds to a finite axion-mass window that can be probed. To ensure continuous coverage of the accessible axion-mass range without gaps, an appropriate step size in $\theta$ must be adopted. Taking differentials of both sides of Eq.~(\ref{rescond}), we obtain 
\begin{equation}
\label{scanrate}
\delta \theta = \frac{\delta m_{a}}{m_{a}} \tan \theta,
\end{equation}
where $\delta m_{a}/m_{a} \approx \Delta \omega/\omega$ defines the width of the mass sensitivity window for a fixed $\theta$. As illustrated in Fig.~\ref{fig5},  the angular step size increases with increasing $\theta$. As evident from the resonance condition, the lower mass range can be probed with smaller $\theta$ values. 
Accordingly, the total number of angular settings, $N_{\theta}$,
required to scan the accessible mass range over the interval
$\theta_{\rm min}\leq\theta\leq\theta_{\rm max}$ can be expressed as
\begin{equation}
\label{nshot}
N_{\theta} = \int_{\rm \theta_{\rm min}}^{\theta_{\rm max}} \frac{d\theta}{\delta \theta}. 
\end{equation}
Inserting the $\delta\theta$ expression from
Eq.~(\ref{scanrate}) into the above integral gives
\begin{equation}
N_{\theta} =
\left(\frac{\delta m_a}{m_a}\right)^{-1}
\ln\left(
\frac{\sin\theta_{\rm max}}{\sin\theta_{\rm min}}
\right).
\end{equation} 
Using the ELI-NP laser parameters, with a carrier wavelength
$\lambda=800~\mathrm{nm}$ and a spectral bandwidth
$\Delta\lambda \approx 56~\mathrm{nm}$, the accessible axion-mass
range $m_{a} \sim 10^{-6}-10~\mathrm{eV}$ can be covered in
approximately $N_{\theta}\simeq197$ angular settings by varying the
beam angle over the interval
$1\,\mu{\rm rad}\leq\theta\leq80^\circ$, with the angular spacing
$\delta\theta$ determined by Eq.~(\ref{scanrate}). The number
$N_{\theta}$ specifies the scan granularity and should not be confused
with the total number of laser shots. With the above scan rate, we now establish the projected sensitivity 
of our proposed experiment in the $g_{a\gamma}-m_{a}$ parameter space. To detect the emitted THG ($3\omega$) photons, one could use superconducting nanowire single-photon detectors (SNSPDs) similar to \cite{Hochberg:2019yo,Chiles:2022jeff}, capable of single photon detection from the mid-infrared to the ultraviolet band. Indeed, SNSPDs have demonstrated ultralow dark count rates ($10^{-6} $ Hz ), which make them ideal to detect rare signal events like the ones in the proposed experiment. Furthermore, their active areas are large enough to collect the focused light ($\gtrsim 0.1 \rm mm^2$ ), enabling 
near-unity detection efficiency and sensitivity to photons from $0.1$  eV to $10$ eV. These properties make SNSPDs well suited to the unique requirements of this proposal's goal to detect $3\omega$-photons with energy $\sim 4.65$ eV (corresponding to $\lambda=800/3$ nm). The instrumental backgrounds include detector dark counts, blackbody
radiation, residual fundamental and harmonic light scattered by the
optical components, and nonlinear emission from residual gas. Thermal
and detector backgrounds can be constrained using laser-off data,
energy discrimination, and coincidence timing with the laser pulses.
Spurious harmonic generation in residual gas can be suppressed by
operating the interaction region under high-vacuum conditions and
monitored by varying the residual-gas pressure.

A qualitatively different contribution originates from the
Euler--Heisenberg nonlinearity of the quantum vacuum. In a
non-collinear laser configuration, photon merging can generate a
coherent $3\omega$ field whose directionality can overlap that of the
axion-induced signal \cite{SundqvistKarbstein2023}. The
Euler--Heisenberg contribution must therefore be treated at the
amplitude level, including its possible interference with the
axion-mediated field.

The two amplitudes nevertheless have different longitudinal coherence
properties. In the monochromatic plane-wave approximation, the
Euler--Heisenberg amplitude accumulated over an interaction length $L$
has the schematic form
\begin{equation}
  \mathcal{A}_{\rm EH}
  \propto
  L\,\exp\left(-\frac{i\Delta k_zL}{2}\right)
  \operatorname{sinc}\left(\frac{\Delta k_zL}{2}\right),
  \label{eq:EH_coherence}
\end{equation}
where $\operatorname{sinc}(x)\equiv\sin(x)/x$ and, for the geometry
considered here,
\begin{equation}
  \Delta k_z =
  k_0\left[
    \sqrt{9-\sin^2\theta}-3\cos\theta
  \right].
  \label{eq:EH_mismatch}
\end{equation}
Using $L\simeq w/\sin\theta$, the leading-order
Euler--Heisenberg amplitude exhibits coherence minima at
\begin{equation}
  w_n =
  \frac{n\lambda\sin\theta}
  {\sqrt{9-\sin^2\theta}-3\cos\theta},
  \qquad n=1,2,\ldots .
  \label{eq:EH_nulls}
\end{equation}
By contrast, the axion-mediated contribution is resonantly phase
matched when $m_a=2\hbar\omega\sin\theta/c^2$ and displays the coherent
buildup derived in Sec.~\ref{sec:thirdharmonic}. This suggests operating
near an Euler--Heisenberg coherence minimum and using neighbouring
values of the beam waist or crossing angle as control configurations.
Angle and polarisation scans provide additional handles for separating
the resonant axion response from instrumental and Standard-Model
contributions.

The exact zeros in Eq.~\eqref{eq:EH_nulls} apply to the idealised
monochromatic plane-wave treatment. Finite pulse duration, focusing,
spectral bandwidth, angular divergence, and detector acceptance will
generally fill these minima. A quantitative analysis will therefore
require propagation of the Euler--Heisenberg field, including its
interference with the axion amplitude, through a realistic beam and
detector model. In the exploratory projections below, we assume that
geometrical optimisation and the use of angular, waist, spectral, and
polarisation control measurements reduce the effective background
expectation in the signal region to below one event. 

Under this assumption, we define the idealised single-photon reach by
\begin{equation}
  \mu_a =
  \epsilon_{\rm det}\,
  \bar N_{\rm THG}^{(\rm res)}=1,
  \label{eq:single_photon_criterion}
\end{equation}
where $\epsilon_{\rm det}$ is the total collection and detection
efficiency and $\bar N_{\rm THG}^{(\rm res)}$ is the axion-mediated
yield integrated over the stated exposure. This condition represents a
single-photon-yield projection rather than a confidence-level
exclusion. For a background-free Poisson process with zero observed events,
$P(0|\mu_a)=e^{-\mu_a}$, giving the one-sided 90\% confidence-level
limit $\mu_a^{90}=-\ln(0.1)=2.30$. The $\mu_a=1$ criterion corresponds
instead to a $1-e^{-1}\simeq63\%$ probability of observing at least one
photon; since $\mu_a\propto g_{a\gamma}^{4}$, using the 90\% limit would
weaken the projected coupling reach by only
$2.30^{1/4}\simeq1.23$, or approximately 23\%.

Using the experimental parameters described above, we obtain the
projected single-photon reach shown in Fig.~\ref{fig6}. Each point of
the figure corresponds to the integrated exposure stated in the caption
at the angular setting that satisfies the axion resonance condition.
The approximately 197 settings estimated above specify the number of
angles required for complete mass coverage.
Note that we can extend the exclusion region to lower axion mass ($m_{a}\lesssim 10^{-5}$ eV) by probing at arbitrarily small angles. In practice,  however, this would necessitate exceedingly small angular step sizes, which is beyond the reach of the proposed setup. Additionally, laboratory-based constraints from other experiments like CROWS \cite{Betz:2013dza}, OSQAR \cite{OSQAR:2007oyv}, ALPS I \cite{Ehret:2010mh,Bahre:2013ywa}, and II \cite{ALPSII:2025eri}, 
PVLAS \cite{PVLAS:2007wzd,Ejlli:2020yhk}, and BaBar \cite{Dolan:2017osp} are also depicted. We see that our proposal is able to probe regions of axion parameter space that are not already excluded by existing LSW/vacuum birefringence experiments. 
Furthermore, we emphasize that due to the cubic scaling on impinging intensity of the laser beams [see Eq.~(\ref{nthgapprox})], greater signal intensity can be achieved by maximizing the impinging intensity. The rapid development of high-intensity laser technology over the past decades suggests that even higher laser intensities (peak power $\sim 100$ PW, see blue line in Fig.~\ref{fig6}) may become available in the future, potentially extending the sensitivity of the proposed search to previously unexplored regions of parameter space. In this direction, higher mass range can also be probed by the collision of X-ray free electron lasers, which offer  higher degree of frequency tuning, allowing easier scanning of parameter space. Future advances in high-power laser technology may enable sensitivity to the QCD axion band in the eV-mass regime, owing to the favorable cubic scaling of the signal-photon yield with laser energy.

\section{Conclusions}

In conclusion, we have proposed a fully laboratory-based strategy to search for axions and axion-like particles through third-harmonic generation induced by two non-collinear, high-intensity laser beams. Starting from axion-modified Maxwell equations, we derived analytically the axion field generated by the pump beams and the resulting THG electromagnetic field. A central feature of the scheme is the resonant enhancement of axion production obtained by tuning the angle between the two beams, which provides direct access to the axion mass without relying on a resonant cavity or on the physical size of the apparatus.

For realistic parameters of the ELI-NP laser system, the resulting THG photon yield can become detectable over a broad region of the $g_{a\gamma}-m_{a}$ parameter space. The finite laser bandwidth allows the accessible mass range to be continuously scanned by varying the collision angle, requiring only a few hundred angular steps over the full interval considered here. The distinctive angular and polarisation dependence of the axion-induced THG signal further provides useful experimental handles for discriminating it from residual optical and detector backgrounds. 
The Euler--Heisenberg nonlinearity provides an additional coherent
$3\omega$ contribution with potentially overlapping directionality.
Its different longitudinal coherence structure suggests that it can be
constrained by operating near its coherence minima and by performing
angle-, waist-, and polarisation-dependent control measurements. A
complete experimental sensitivity analysis will nevertheless require a
finite-pulse calculation of this contribution and of its interference
with the axion-mediated amplitude.

Our results, therefore, identify NL frequency conversion as a complementary route for laboratory axion searches, exploiting the extreme field strengths, short pulse durations, and geometric tunability available at modern petawatt laser facilities. The strong (cubic) scaling of the signal with the impinging laser intensity 
also makes the proposed approach particularly promising in view of future developments toward higher-power laser systems, which could significantly extend the accessible parameter space.  

\section*{Acknowledgements}
 The authors thank Zurab Berezhiani, Giacomo Marocco and Fabrizio Nesti for useful discussions. 
 The work of LDL is supported by the Italian Ministry of University and Research (MUR) via the FIS2 Consolidator Grant project FIS-2023-02106 -- QAXION (CUP: I53C25001880001).

 
\appendix
\section{Derivation of the total axion solution in Eq.~(\ref{phisol})}
\label{appA}
Since the axion-induced THG fields are second-order in the coupling constant, i.e.,  $E_{\rm THG} \propto g_{a\gamma}^2$, it is enough to calculate the right hand side of Eq.~(\ref{KGphi}) at the leading order $ \mathcal{O}(g_{a\gamma}^0)$, i.e., Re[$\textbf{E} \cdot \textbf{B}] \simeq \frac{1}{2} \text{Re}\big[  \textbf{E}_0 \cdot \textbf{B}_0^{*} + \textbf{E}_0 \cdot \textbf{B}_0 e^{-2i\omega t} \big]$, where $\textbf{E}_0=E_+ \hat{\bf n}_+ e^{i\textbf{k}_+ \cdot \textbf{r} } + E_-\hat{\bf n}_- e^{i\textbf{k}_- \cdot \textbf{r} }$ and $\textbf{B}_0= \omega^{-1} \Big[ E_+ \textbf{k}_+\times \hat{\bf n}_+ e^{i\textbf{k}_+ \cdot \textbf{r} } + E_-\textbf{k}_-\times\hat{\bf n}_- e^{i\textbf{k}_- \cdot \textbf{r} }   \Big]$. Therefore, Eq.~(\ref{KGphi}) reduces to 
\begin{align} 
\label{edotb} 
& \left(\frac{1}{c^2}\frac{\partial^2}{\partial t^2} - \nabla^2 +  \frac{m_{a}^2 c^2}{\hbar^2}\right) a =  \\ \nonumber & \frac{g_{a\gamma} \sin^2\theta}{\mu_0 c^2} \text{Re} \bigg[E_+E_-^* \left( \alpha_{\rm TE}^{(-)*} \alpha_{\rm TM}^{(+)} +  \alpha_{\rm TM}^{(-)*} \alpha_{\rm TE}^{(+)} \right)  e^{i\left(\textbf{k}_+ - \textbf{k}_- \right) \cdot \textbf{r}} \\ \nonumber & + E_+E_- \left( \alpha_{\rm TE}^{(+)} \alpha_{\rm TM}^{(-)} +  \alpha_{\rm TM}^{(+)} \alpha_{\rm TE}^{(-)} \right)  e^{i\left(\textbf{k}_+ + \textbf{k}_- \right) \cdot \textbf{r}} e^{-2i\omega t}\bigg] .
\end{align}
Inserting the Ansatz $a(\textbf{r},t)=\text{Re}\big[a_0(\textbf{r})+a_2(\textbf{r})e^{-2i\omega t} \big]$ into the above equation and separating out the static ($e^{0}$) and second harmonic ($e^{-2i\omega t}$) parts we obtain the following PDEs 
\begin{subequations}
\begin{align}  
\label{eq:a0}
& \left( - \nabla^2 +  \frac{m_{a}^2 c^2}{\hbar^2}\right) a_0 =  \\ \nonumber & \frac{g_{a\gamma} \sin^2\theta}{\mu_0 c^2}E_+E_-^* \left( \alpha_{\rm TE}^{(-)*} \alpha_{\rm TM}^{(+)} +  \alpha_{\rm TM}^{(-)*} \alpha_{\rm TE}^{(+)} \right)    e^{i\left(\textbf{k}_+ - \textbf{k}_- \right) \cdot \textbf{r}},  \\
\label{eq:a2}
& \left( -\frac{4\omega^2}{c^2}- \nabla^2 +  \frac{m_{a}^2 c^2}{\hbar^2}\right) a_2 = \\ \nonumber &  \frac{g_{a\gamma} \sin^2\theta}{\mu_0 c^2}  E_+E_- \left( \alpha_{\rm TE}^{(+)} \alpha_{\rm TM}^{(-)} +  \alpha_{\rm TM}^{(+)} \alpha_{\rm TE}^{(-)} \right)    e^{i\left(\textbf{k}_+ + \textbf{k}_- \right) \cdot \textbf{r}}.
\end{align}
\end{subequations}
The particular solutions of the above PDEs can readily be obtained as 
\begin{align}  
a_0^{\rm (p)} &= \frac{g_{a\gamma} \sin^2 \theta \Big[ E_+E_-^{*}  \left( \alpha_{\rm TE}^{(-)*} \alpha_{\rm TM}^{(+)} + \alpha_{\rm TE}^{(+)}\alpha_{\rm TM}^{(-)*}  \right)\Big]}{\mu_0 c^2 \Big[ \lvert \textbf{k}_+ - \textbf{k}_- \rvert^2 +  \frac{m_{a}^2c^2}{\hbar^2} \Big]} \nonumber \\ 
&\times e^{i \left( \textbf{k}_+ - \textbf{k}_-\right) \cdot \textbf{r}} , \\
a_2^{\rm (p)} &= \frac{g_{a\gamma} \sin^2 \theta \Big[ E_+E_- \left( \alpha_{\rm TE}^{(-)} \alpha_{\rm TM}^{(+)} + \alpha_{\rm TE}^{(+)}\alpha_{\rm TM}^{(-)}  \right)\Big]}{\mu_0 c^2 \Big[ -\frac{4\omega^2}{c^2} + \lvert  \textbf{k}_+ + \textbf{k}_- \rvert^2 +  \frac{m_{a}^2c^2}{\hbar^2} \Big]} \nonumber \\ &\times e^{i \left( \textbf{k}_+ + \textbf{k}_-\right) \cdot \textbf{r}} . 
\end{align}
The total axion field, solution of Eq.~(\ref{phisol}), is obtained by adding the homogeneous solutions to the particular one obtained above. The homogeneous solutions $(a_0^{\rm (h)}, a_2^{\rm (h)} )$ are readily obtained by setting the RHS of Eqs.~(\ref{eq:a0}, \ref{eq:a2}) to zero, and by applying the BC of the vanishing total axion field at $z=0$ for every $x$ and $t$, obtaining $a_0^{\rm (h)}(z=0)=-a_0^{\rm (p)}(z=0)$ and $a_2^{\rm (h)}(z=0)=-a_2^{\rm (p)}(z=0)$.

\vspace{10ex}

\section{Derivation of third harmonic field expressions in Eq.~(\ref{ebres})}
\label{appB}

To obtain the field expressions in Eq.~(\ref{ebres}) at the resonance condition $\sin \theta = m_{a}c^2/2\hbar \omega$, first we calculate  the field expressions at arbitrary $\theta$. This is accomplished by decoupling the MMEs [Eq.~(\ref{MEaxion})] in terms of the perturbed $(\textbf{E}_{\rm THG})$ and unperturbed  $(\textbf{E}_{0})$ contributions, obtaining
\begin{eqnarray} 
\label{ME3omg}
&& \nabla \cdot \textbf{E}_{\rm THG} = - \frac{g_{a\gamma}c}{2} \textbf{B}_0 \cdot \nabla a_2 ,\\ \nonumber
&& \nabla \cdot \textbf{B}_{\rm THG}  = 0 , \\   \nonumber
&& \nabla \times \textbf{E}_{\rm THG}  = 3i\omega  \textbf{B}_{\rm THG} , \\   \nonumber
&& \nabla \times \textbf{B}_{\rm THG} + \frac{3i\omega}{c^2}  \textbf{E}_{\rm THG} = - \frac{g_{a\gamma}}{2c} \Big[2i\omega\textbf{B}_0 a_2 +  \textbf{E}_0 \times \nabla a_2   \Big].
\end{eqnarray}
Inserting the expression of $a_2$  into Eq.~(\ref{ME3omg}) and after some simplifications, the above sets of equations can be reduced to the single inhomogeneous Helmholtz equation 
\begin{widetext}
\begin{align} 
\label{ME3hemholtz1} 
\big[\nabla^2 + 9k_0^2 \big]  \textbf{E}_{\rm THG} &= k_0^2   \frac{g_{a\gamma}^2 \sin^2 \theta \Big[ E_+E_- \left( \alpha_{\rm TE}^{(-)} \alpha_{\rm TM}^{(+)} + \alpha_{\rm TE}^{(+)}\alpha_{\rm TM}^{(-)}  \right)\Big]}{\mu_0 c^2 \Big[ -\frac{4\omega^2}{c^2} + \lvert  \textbf{k}_+ + \textbf{k}_- \rvert^2 +  \frac{m_a^2c^2}{\hbar^2} \Big]} \nonumber \\
&\times \begin{bmatrix} E_+    \sin^2 \theta \left( \cos \theta  \alpha_{\rm TE}^{(+)} \hat{\textbf{x}} -3 \alpha_{\rm TM}^{(+)} \hat{\textbf{y}} - 3 \alpha_{\rm TE}^{(+)} \sin \theta \hat{\textbf{z}}  \right)   e^{i\left(2\textbf{k}_+ + \textbf{k}_- \right) \cdot \textbf{r}} \\ 
+E_+ \biggl\{  \alpha_{\rm TE}^{(+)} \left(3\gamma_{a}-\gamma_{a} \sin^2 \theta - 3\cos \theta \right) \hat{\textbf{x}} + 3  \alpha_{\rm TM}^{(+)} \left(1- \gamma_{a} \cos \theta \right)  \hat{\textbf{y}} \\ +   \alpha_{\rm TE}^{(+)} \sin \theta \left(3-\gamma_{a} \cos \theta - 2\gamma_{a}^2 \right) \hat{\textbf{z}} \biggl\} e^{i \textbf{k}_+ \cdot \textbf{r} + 2ik_0 \gamma_{a} z} 
\\ + E_-   \sin^2 \theta \left( \cos \theta  \alpha_{\rm TE}^{(-)} \hat{\textbf{x}} -3 \alpha_{\rm TM}^{(-)} \hat{\textbf{y}} + 3 \alpha_{\rm TE}^{(-)} \sin \theta \hat{\textbf{z}}  \right)    e^{i\left(2\textbf{k}_- + \textbf{k}_+ \right) \cdot \textbf{r}} \\
+E_- \biggl\{  \alpha_{\rm TE}^{(-)} \left(3\gamma_{a}-\gamma_{a} \sin^2 \theta - 3\cos \theta \right) \hat{\textbf{x}} + 3  \alpha_{\rm TM}^{(-)} \left(1- \gamma_{a} \cos \theta \right)  \hat{\textbf{y}} \\ -   \alpha_{\rm TE}^{(-)} \sin \theta \left(3-\gamma_{a} \cos \theta - 2\gamma_{a}^2 \right) \hat{\textbf{z}} \biggl\} e^{i \textbf{k}_- \cdot \textbf{r} + 2ik_0 \gamma_{a} z}
\end{bmatrix},
\end{align}
where $k_0=\omega/c$ and $\gamma_{a}=\sqrt{1-m_{a}^2 c^4/4\hbar^2\omega^2}$, which readily leads to the particular solution
\begin{align} 
\label{Ep3omg} 
 \textbf{E}_{\rm THG}^{\rm (p)} &= \frac{1}{8}  \frac{g_{a\gamma}^2 \sin^2 \theta \Big[ E_+E_- \left( \alpha_{\rm TE}^{(-)} \alpha_{\rm TM}^{(+)} + \alpha_{\rm TE}^{(+)}\alpha_{\rm TM}^{(-)}  \right)\Big]}{\mu_0 c^2 \Big[ -\frac{4\omega^2}{c^2} + \lvert  \textbf{k}_+ + \textbf{k}_- \rvert^2 +  \frac{m_{a}^2c^2}{\hbar^2} \Big]} \nonumber \\ 
 &\times \begin{bmatrix} E_+    \sin^2 \theta \left( \cos \theta  \alpha_{\rm TE}^{(+)} \hat{\textbf{x}} -3 \alpha_{\rm TM}^{(+)} \hat{\textbf{y}} - 3 \alpha_{\rm TE}^{(+)} \sin \theta \hat{\textbf{z}}  \right)   e^{i\left(2\textbf{k}_+ + \textbf{k}_- \right) \cdot \textbf{r}} \\ 
+\frac{2E_+ }{\left(2-\gamma_{a} \cos\theta - \gamma_{a}^2\right)}\biggl\{  \alpha_{\rm TE}^{(+)} \left(3\gamma_{a}-\gamma_{a} \sin^2 \theta - 3\cos \theta \right) \hat{\textbf{x}} + 3  \alpha_{\rm TM}^{(+)} \left(1- \gamma_{a} \cos \theta \right)  \hat{\textbf{y}} \\ +   \alpha_{\rm TE}^{(+)} \sin \theta \left(3-\gamma_{a} \cos \theta - 2\gamma_{a}^2 \right) \hat{\textbf{z}} \biggl\} e^{i \textbf{k}_+ \cdot \textbf{r} + 2ik_0 \gamma_{a} z} 
\\ + E_-   \sin^2 \theta \left( \cos \theta  \alpha_{\rm TE}^{(-)} \hat{\textbf{x}} -3 \alpha_{\rm TM}^{(-)} \hat{\textbf{y}} + 3 \alpha_{\rm TE}^{(-)} \sin \theta \hat{\textbf{z}}  \right)    e^{i\left(2\textbf{k}_- + \textbf{k}_+ \right) \cdot \textbf{r}} \\
+\frac{2E_- }{\left(2-\gamma_{a} \cos\theta - \gamma_{a}^2\right)}\biggl\{  \alpha_{\rm TE}^{(-)} \left(3\gamma_{a}-\gamma_{a} \sin^2 \theta - 3\cos \theta \right) \hat{\textbf{x}} + 3  \alpha_{\rm TM}^{(-)} \left(1- \gamma_{a} \cos \theta \right)  \hat{\textbf{y}} \\ -   \alpha_{\rm TE}^{(-)} \sin \theta \left(3-\gamma_{a} \cos \theta - 2\gamma_{a}^2 \right) \hat{\textbf{z}} \biggl\} e^{i \textbf{k}_- \cdot \textbf{r} + 2ik_0 \gamma_{a} z}
\end{bmatrix}.
\end{align}
\end{widetext}
The particular solution of the corresponding magnetic field is then calculated as $\textbf{B}_{\rm THG}^{\rm (p)} = -\frac{i}{3\omega}\nabla \times \textbf{E}_{\rm THG}^{\rm (p)}$. To obtain the total solution, we add the homogeneous solution to the particular solution $\textbf{B}_{\rm THG}=\textbf{B}_{\rm THG}^{\rm (p)}+\textbf{B}_{\rm THG}^{\rm (h)}$, where the homogenous solution satisfies $(\nabla^2 + 9k_0^2) \textbf{B}_{\rm THG}^{\rm (h)}=0$. The coefficients of the homogeneous solution is then determined from the boundary conditions:
\begin{eqnarray}
\label{bcond}
B_{\rm THG}^x(z=0)=0, \\ B_{\rm THG}^y(z=0)=0,  \\ \nabla \cdot \textbf{B}_{\rm THG}=0.
\end{eqnarray}
Note that with the above choice one automatically obtains a vanishing the $z$-component of the electric field at $z=0$, i.e.,  $E_{\rm THG}^z(z=0)=0$.  Consequently, the axion-induced third harmonic intensity also vanishes at $z=0$. The resonance expressions of Eq.~(\ref{ebres}) can then be obtained by taking a Taylor expansion of the field about the resonance condition, inserting it $\sin\theta = (m_{a}c^2/2 \hbar \omega)  +y$ in  Eq.~(\ref{Ep3omg}) and taking the limit $y \rightarrow 0$. 

\bibliographystyle{apsrev4-1.bst}
\bibliography{bibliography}

\end{document}